\documentclass[a4paper]{jpconf}
\usepackage{graphicx,amssymb}
\begin{document}
\newcommand{\be}{\begin{equation}}
\newcommand{\ee}{\end{equation}}
\newcommand{\beq}{\begin{eqnarray}}
\newcommand{\eeq}{\end{eqnarray}}
\def\lsim{\hbox{ \raise.35ex\rlap{$<$}\lower.6ex\hbox{$\sim$}\ }}
\def\gsim{\hbox{ \raise.35ex\rlap{$>$}\lower.6ex\hbox{$\sim$}\ }} 

\title{Lattice refinement in loop quantum cosmology}

\author{Mairi Sakellariadou}

\address{Department of Physics, King's College London, University of
  London, Strand WC2R 2LS, London, U.K.}
          
\ead{mairi.sakellariadou@kcl.ac.uk}

\begin{abstract}
Lattice refinement in LQC, its meaning and its necessity are
discussed. The r\^ole of lattice refinement for the realisation of a
successful inflationary model is explicitly shown. A simple and
effective numerical technique to solve the constraint equation for any
choice of lattice refinement model is briefly
illustrated. Phenomenological and consistency requirements leading to
a particular choice of lattice refinement model are presented, while
it is subsequently proved that only this choice of lattice refinement
leads to a unique factor ordering in the Wheeler-De Witt equation,
which is the continuum limit of LQC.
       
\end{abstract}

\section{Introduction}

Loop Quantum Gravity (LQG)~\cite{rovelli04,thiemann07} is one of the
most promising candidate theories for describing the quantum degrees
of freedom of the gravitational field. Quantum gravity, combining
consistently quantum mechanics and general relativity, is essential
when curvature becomes large, as for example in the early stages of
the evolution of our universe. LQG is a non-perturbative and
background independent\footnote{Background independence means that
  quatisation is achieved in the absence of a metric other that the
  physical one determined by the densitised triad. At the classical
  level, the fact that the laws of physics are background independent
  is mathematically expressed by the Einstein equations being
  four-diffeomorphism covariant.}  canonical quantisation of
General Relativity (GR) in four space-time dimensions. Considering a
fundamental theory of quantum gravity and following the main concept
of GR, namely that gravity is indeed geometry, it follows that there
is no background metric, there is only a manifold, while geometry and
matter should both have a quantum mechanical origin. This indeed
differentiates LQG from other approaches of quantising gravity, which
have been developed in the framework of particle physics\footnote{The
  only known consistent perturbative approach to quantum gravity is
  string theory, a theory aiming at unifying all interactions.}.
While the full theory is not yet complete, LQG already has a number of
successes, such as the construction of quantum geometry, the
prediction of Planck discreteness in geometric operators, as well as a
quantum accounting for black hole entropy on the horizon.

The application of ideas and mathematical methods of the full LQG
theory to the cosmological sector --- the dynamical variables are
reduced to first homogeneity and then possibly also to isotropic
models --- led to Loop Quantum Cosmology (LQC)~\cite{bojowald05},
which is not a field theory\footnote{The quantised Hamiltonian is
  given in terms of symmetry reduced variables, so there is only a
  finite number of degrees of freedom.}.  LQC, which gains a
constantly increasing interest from the scientific community and has
recently made significant progress, differs from other attempts of
quantising gravity in the sense that it gets its input from the full
theory of LQG.  In short, it is a cosmological mini-superspace model
quantised with methods of the full LQG theory.  Considering a flat
isotropic model within LQC, the extrinsic curvature scale $k=\dot
a=\sqrt{8\pi Ga^2\Lambda/3}$ ($a$ stands for the scale factor and
$\Lambda$ denotes a positive cosmological constant) appears in the
holonomies in such a way that only $e^{i\alpha k}$, with
$\alpha\in{\bf R}$, can be represented as operators, $k$ itself
cannot~\cite{abl}. The parameter $\alpha$ is related to the edge
length used in holonomies, which are playing the r\^ole of the basic
operators and they imply that the Hamiltonian constraint --- a
Hamiltonian density which is constrained to vanish by the equations
of motion --- is quantised to a second order difference equation,
instead of the second order differential equation of the Wheeler-De
Witt (WDW) approach to quantum cosmology.

Following the approach of LQC is useful in two
ways. Firstly, it allows us to get some useful insight about open
issues of the full LQG theory, and secondly, by using symmetry
reduction of the infinite dimensional phase space of LQG, the theory
becomes often tractable leading to satisfactory answers about various
interesting physical questions.  The discreteness of spatial geometry,
a key element of the full theory, leads to successes in LQC which do
not hold on the WDW approach.  In particular, it has been shown that
classical big bang~\cite{bojowald01} and black hole
singularities~\cite{ashtekar-etal00} are removed in LQC, in a well
defined manner.

In LQC, the evolution of the universe is divided into three distinct
phases depending on the value of the scales probed by the universe
itself. They are the following:
\begin{itemize}
\item{\sl The discrete quantum phase}: Very close to the Planck scale,
  the concept of space-time has no meaning and one should consider the
  full quantum gravity theory. Applying LQC during this phase, leads to a
  finite bounded spectrum for eigenvalues of inverse powers of the
  three-volume density, called the geometrical density.
\item{\sl The intermediate phase}: As the volume of the universe
  increases with time, the universe enters a semi-classical
  phase. More precisely, for lengths above $L_{\rm Pl}\equiv
  {\sqrt\gamma} l_{\rm Pl}$ ($\gamma\approx 0.2375$ is the
  Barbero-Immirzi parameter --- a constant ambiguity parameter, whose
  value is fixed by black hole entropy calculations --- and $l_{\rm
    Pl}$ is the Planck length with $l_{\rm Pl}^2=(8\pi G)^{-1}$) the
  space-time can be approximated by a continuous manifold and the
  equations of motion take a continuous form. This intermediate phase
  is characterised by a second scale $L_\star$ with
  $L_\star\equiv\sqrt{(\gamma J \mu_0)/3}l_{\rm Pl}$, below which the
  geometrical density is significantly different from its classical
  form\footnote{The half-integer $J$ labels the ambiguity in choosing
    the representation in which the matter part of the Hamiltonian
    constraint for a scalar field is quantised. The length parameter
    $\mu_0$ is related to the underlying discrete structure.}. For
  scales below $L_\star$ quantum corrections cannot be neglected. This
  phase is the most relevant one regarding phenomenological
  consequences of LQC.
\item{\sl The classical phase}: At later times, and
therefore large scales, the universe enters the full classical phase
and standard cosmology becomes valid.
\end{itemize}

LQG/LQC is formulated in terms of SU(2) holonomies of the connection
and triads\footnote{The connection in LQG determines the parallel
  transport of chiral fermions, mathematically represented by spinors,
  while the conjugate momenta can be interpreted as spatial triads
  (i.e., {\sl square-roots} of the metric of the 3-dimensional
  space). }~\cite{dv}.  To obtain a quantum constraint we introduce an
operator representing the curvature of the gravitational connection.
A feature of LQC, which is a direct consequence of LQG, is that while
there are well-defined analogs of holonomies, there is no operator
corresponding to the connection; one defines a curvature operator in
terms of holonomies. In the classical level, curvature can be
expressed as a limit of the holonomies around a loop as the area
enclosed by the loop shrinks to zero. However, in quantum geometry one
cannot continuously shrink a loop to zero area, since the eigenvalues
of the area operator are discrete, implying that there is a smallest
non-zero area eigenvalue, called the {\sl area gap}~\cite{almmt,al}.
In a canonical quantisation scheme, as the one followed in this
approach, one first writes the action of GR into a Hamiltonian
formulation and then one quantises this classical Hamiltonian.  In the
{\sl old} quantisation, the quantised holonomies were taken to be
shift operators with a fixed magnitude. However, it was
shown~\cite{Rosen:2006bga,Bojowald:2007ra} that this approach leads to
unavoidable instabilities in the continuum semi-classical limit, where
the WDW wave-function becomes a good approximation to the difference
equation of LQC. For a large semi-classical universe, the WDW
wave-function would be oscillating on scales of the order
$(a\sqrt\Lambda)^{-1}$. As the universe expands, this scale becomes
eventually smaller than the discreteness scale of the difference
equation of LQC, implying that discreteness of spatial geometry would
become apparent in the behaviour of wave-functions describing a
classical universe.  In the underlying LQG theory, the contributions
to the discrete Hamiltonian operator depend on the state which
describes the universe. As the universe expands, the full Hamiltonian
constraint operator creates new vertices of a lattice state, in
addition to changing their edge labels. As the extrinsic curvature
scale $k$ increases with increasing volume, the corresponding $\alpha$
decreases since the lattice is being continuously refined. Thus, in
the context of LQC one has a refinement of the discrete
lattice\footnote{The parameter, denoted earlier by $\mu_0$, which
  appears in the regularisation of the Hamiltonian constraint will no
  longer be constant within the lattice refinement context.}. One
can choose such a lattice refinement model, so that the increase in
extrinsic curvature scale $k$ can be balanced by the decrease of
$\alpha$ such that $\alpha k$ remains small and semi-classical
behaviour is achieved for any macroscopic volume even in the presence
of a positive cosmological constant.

Lattice refinement is also required from phenomenological
reasons~\cite{Bojowald:2007ra,Nelson:2007um}. For example, as we will
discuss, lattice refinement renders a successful inflationary
era\footnote{It was hoped that LQC would help to overcome the extreme
  fine tuning necessary to achieve successful inflation in
  GR~\cite{cs1,cs2,gt}. However, it has been shown~\cite{gns} that
  semi-classical corrections are insufficient to alleviate this
  difficulty. Certainly, inflation could be generic in the deep
  quantum regime.} more natural~\cite{Nelson:2007um}. The effect of
lattice refinement has been modelled and the elimination of the
instabilities in the continuum era has been explicitly shown.

The {\sl correct} lattice refinement model should be obtained from the
full LQG theory. In principle, one should use the full Hamiltonian
constraint and find the way that its action balances the creation of
new vertices as the volume increases. Instead, phenomenological
arguments have been used, where the choice of the lattice refinement
model is constrained by the form of the matter Hamiltonian~\cite{ns2}.
In particular, we have shown~\cite{ns2} that for LQC to generically
support inflation, and other matter fields, without the onset of
large scale quantum gravity corrections, one should adopt a particular
model of lattice refinement. This choice has been then
found~\cite{gs08} to be the only one, for which physical quantities are
independent of the choice of the elementary cell used to regulate the
spatial integrations.  Amazingly enough, this is exactly the choice
required for the uniqueness of the factor ordering of the Wheeler-De
Witt equation~\cite{ns4}.

Lattice refinement leads to new dynamical difference equations which,
in general, do not have a uniform step-size making their study quite
involved, particularly for anisotropic cases, as for example for
Bianchi models or black hole interiors. Numerical techniques have been
developed~\cite{nt1,ns3} to address this issue.

Our primary interest here concerns cosmological predictions of quantum
gravity, and therefore we only focus on questions dealing with the
early universe and the initial conditions. A quantum theory of gravity
is expected, on the one hand, to cure the classical singularities of GR
and, on the other hand, either to provide the conditions suitable for the
onset of inflation, or to suggest an alternative scenario for alleviating
the strong fine-tuning of the standard cosmological model.

\section{Elements of LQG/LQC}

LQG, as well as LQC, are both based on a Hamiltonian formulation of GR
with basic variables an SU(2) valued connection and the conjugate
momentum variable which is a densitised triad\footnote{The variables
  $E^a_i$ and $A^i_a$ were introduced by Barbero~\cite{barbero95} as
  an alternative to the complex Ashtekar~\cite{ashtekar86}
  variables. Both real and complex connections have been used for
  canonical gravity. The complex connection has SL(2,C) as gauge
  group, while the real connection has SU(2) as gauge
  group. Mathematical techniques can only cope with a quantum theory
  based on SU(2).}, a derivative operator quantised in the full LQG
theory in the form of fluxes. By using connection-triad variables,
arising from a canonical transformation of Arnowitt-Desner-Misner
(ADM) variables, we make an analogy with gauge theories; this will be
helpful when dealing with quantisation issues.

The densitised triad carries information about the spatial geometry,
encoded in the three-metric; the connection carries information about
the spatial curvature, in the form of the spin-connection, and the
extrinsic curvature. More precisely, the densitised triad, $E_i^a$, is
related to the three-metric, $q_{ab}$, by
\be 
E_i^a=\sqrt{|q|}e^a_i~, 
\ee 
where $e^a_i$ is a physical triad, dual ($e^a_ie^j_a=\delta^j_i$) to
the co-triad, $e^j_a$, and satisfying
\be 
q_{ab}=e^i_ae^j_b\delta_{ij}~. 
\ee 
Note that $i$ refers to the Lie algebra
  index and $a$ is a spatial index with $a,i=1,2,3$.

The connection, $A_a^i$, can be related to the ADM variables as
\be
A_a^i=\Gamma_a^i+\gamma K_a^i~,
\ee
where $\Gamma_a^i=-(1/2)\epsilon^{ijk}e_j^b(2\partial_{[a}e^k_{b]}
+e^c_ke^l_a\partial_ce^l_b)$ stands for the spin-connection compatible
with the co-triad, $\gamma$ is the Barbero-Immirzi parameter ---
classically it has no physical consequence, while at the quantum level
it plays a r\^ole in the level spacing of discrete geometric
eigenvalues --- and $K_a^i$ stands for the extrinsic curvature
one-form $K_a^i=e^{bi}K_{ab}$ (with $K_{ab}$ the extrinsic curvature),
it is the Lie derivative of $q$ with respect to the normal vector to
the spatial slice, $K^i_a=({\cal L}_{\vec n}q_{ab})\delta^{ij}e^b_j$.

The Poisson bracket of the densitised triad and the connection reads
\be
\{A_a^i(x),E_j^b(y)\}= \kappa\gamma\delta^b_a\delta^i_j\delta^3(x,y)~,
\ee
where $\kappa\equiv 8\pi G$.

The Hamiltonian for GR is given by the sum of constraints, with the
scalar constraint
\be
C_{\rm GR}={1\over 2\kappa}\int_\Sigma {\rm d}^3x N(x)
\left[{E^a_iE^b_j\over\sqrt{|{\rm det}E|}}\epsilon^{ij}_kF^k_{[ab]}-2
(1+\gamma^2){E^a_iE^b_j\over\sqrt{|{\rm det}E|}}K^i_{[a}K^j_{b]}\right]~,
\ee
being the most important one. Note that
$F^k_{[ab]}=2\partial_{[a}A^k_{b]}+\epsilon^k_{\ ij}A^i_{[a}A^j_{b]}$
  denotes the curvature two-form of the connection, and $N$ stands for
  the lapse function.

Before proceeding, let us note that for any quantisation scheme based
on a Hamiltonian framework, as for example LQC, or an action
principle, as for example path integral approaches, for the
homogeneous flat ($k=0$) model, one should regularise the divergences
which appear due to the homogeneity as the action and Hamiltonian are
integrated over spatial hyper-surfaces.  To do so, the spatial
homogeneity and Hamiltonian are restricted to a fiducial
cell\footnote{Note that $\mu_0$, which we will discuss later is the
  scale of the finite fiducial cell that spatial integration is
  restricted to, so as to remove the divergences that occur in
  non-compact topologies.}, with finite volume $V_0=\int{\rm
  d}^3x\sqrt{|^0q|}$~\cite{abl}.

The expressions can be simplified a lot by restricting the
analysis to homogeneous and isotropic geometries:
\be
q=[a(t)]^2\delta_{ij}\ ^0\omega^i_a\ ^0\omega^j_b{\rm d}x^a{\rm d}x^b~,
\ee
where the one-forms satisfy $\partial_{[a}\ ^0\omega^i_{b]}=0$, leading
to
\be
q=[a(t)]^2[(1-kr^2)^{-1}{\rm d}r^2+r^2{\rm d}\Omega^2]~.
\ee
The symmetry reduced variables $E, A$ are given by
\be
A^i_a=V_0^{-1/3} c \ ^0\omega_a^i\ \ \ \ ,
\ \ \ \ E_i^a=pV_0^{-2/3}|{\rm det}\ ^0\omega|\ ^0e^a_i~,
\ee
where the vector fields $^0e^a_i$ are dual to the 1-forms:
$^0\omega^i_a\ ^0e^a_j=\delta^i_j.$

Concentrating on isotropic cosmological models with constant spatial
curvature, we will first consider symmetric connections and triads,
and we will then insert them into the full action leading to a
symmetry reduced action. The symmetric connections and triads can be
decomposed using basis one-forms and vector fields obtained by Bianchi
models. One can show that the actions lead to the correct equations of
motion of GR, implying that they are equivalent to the
Einstein-Hilbert action on metric variables.

The loop quantisation of the flat FLRW scalar constraint changes the
curvature 2-form $F$, its $ab$ component reads~\cite{aps}:
\be  F_{ab}^k =  -2\ {\lim}_{Ar_\Box
  \rightarrow 0} {\rm Tr}
\left(\frac{h^{(\bar\mu)}_{\Box_{ij}}-1 }{\bar\mu^2V_0^{2/3}} \right)
\tau^k \ ^0\omega^i_a \ ^0\omega^j_b~ ;
\ee
$Ar_\Box$ is the area of the square $\Box_{ij}$ in the
$(i,j)$-plane swept by a face of the elementary cell, the holonomy
$h^{(\mu_0)}_{\Box_{ij}}$ around the square $\Box_{ij}$ is the product
    of holonomies along the four edges of $\Box_{ij}$, and $\bar\mu$ is
    $\bar\mu=\sqrt{\Delta}{/|p|^{1/2}}$, with $\Delta$ the eigenvalue
    of the area operator.

In the case of a spatially flat background, derived from the Bianchi
I model, the isotropic connection can be expressed in terms of the
dynamical component of the connection ${\tilde c}(t)$ as 
\be
A_a^i={\tilde c}(t)\omega_a^i~, 
\ee 
with $\omega_a^i$ a basis of left-invariant one-forms $\omega_a^i={\rm
  d}x^i$.  The densitised triad can be decomposed using the Bianchi I
basis vector fields $X_i^a=\delta_i^a$ as 
\be 
E_i^a=\sqrt{^0q}{\tilde p}(t)X_i^a~, 
\ee 
where $^0q$ stands for the determinant of the fiducial background
metric, 
\be 
^0q_{ab}=\omega^i_a\omega_{bi}~, 
\ee 
and ${\tilde p}(t)$ denotes the remaining dynamical quantity after
  symmetry reduction.

In terms of the metric variables with three-metric
$q_{ab}=a^2\omega_a^i\omega_{bi}$, the dynamical quantity is just the
scale factor $a(t)$. Given that the Bianchi I basis vectors are
$X_i^a=\delta^a_i$,
\be
|\tilde{p}|=a^2~,
\ee
where the absolute value is taken because the triad has an
orientation.  Since the basis vector fields are spatially constant in
the spatially flat model, the connection component is
\be
{\tilde c}={\rm sgn}({\tilde p})\gamma \frac{\dot a}{N}~.
\ee
Note that in what follows, the lapse function which is a constant
due to spatial homogeneity, will be set equal to 1.  Thus, GR can
be formulated as a gauge theory in Ashtekar variables.

The canonical variables ${\tilde c}, {\tilde p}$ are related through
\be \{{\tilde c}, {\tilde p}\}=\frac{\kappa\gamma}{3}V_0~, \ee where
$V_0$ the volume of the elementary cell adapted to the fiducial triad.

Defining the triad component $p$, determining the physical volume of
the fiducial cell, and the connection component $c$, determining the
rate of change of the physical edge length of the fiducial cell, as
\be
p=V_0^{2/3}{\tilde p}~~~~~,~~~~~c=V_0^{1/3}{\tilde c}~,
\ee 
respectively, we obtain
\be
\{c,p\}=\frac{\kappa\gamma}{3}~,
\ee
independent of the volume $V_0$ of the fiducial cell.

To quantise, we follow the approach used in the full LQG
theory. The metric itself is a physical field which must be quantised;
it cannot be considered as a fixed background.  Thus, to quantise
gravity we use gauge theory variables to define holonomies of the
connection along a given edge
\be
h_e(A)={\cal P}\exp\int{\rm d}s\dot\gamma^\mu(s)A_\mu^i(\gamma(s))\tau_i~,
\ee
where ${\cal P}$ indicates a path ordering of the exponential,
$\gamma^\mu$ is a vector tangent to the edge and
$\tau_i=-i\sigma_i/2$, with $\sigma_i$ the Pauli spin matrices, and
fluxes of a triad along an $S$ surface
\be
E(S,f)=\int_S\epsilon_{abc}E^{ci}f_i{\rm d}x^a{\rm d}x^b~,
\ee
with $f_i$ an SU(2) valued test function.  Note that even though
these variables appear rather artificial, as presented here in the
context of LQC, they nevertheless arise naturally within the full LQG
theory.

Thus, the basic configuration variables in LQC are holonomies of the
connection 
\be 
h_i^{(\mu_0)}(A)=\cos\left({\mu_0c\over 2}\right){\bf 1}+2\sin\left({\mu_0c\over
  2}\right)\tau_i ~,
\ee
along a line segment $\mu_0\ ^0e^a_i$
and the flux of the triad
$$F_S(E,f)\propto p~;$$ 
the basic momentum variable is the triad component $p$.  Note that
${\bf 1}$ is the identity $2\times 2$ matrix and $\tau_i=-i\sigma_i/2$
is a basis in the Lie algebra SU(2) satisfying the relation
$$\tau_i\tau_j=(1/2)\epsilon_{ijk}\tau^k-(1/4)\delta_{ij}~.$$

Let us first review the {\sl old quantisation} procedure. We follow
the same approach as in LQG. We thus take $e^{i\mu_0c/2}$ (with
$\mu_0$ an arbitrary real number) and $p$, as the elementary classical
variables, which have well-defined analogues~\cite{abl}.  Using the
Dirac bra-ket notation and setting $e^{i\mu_0c/2}=\langle
c|\mu\rangle$, the action of the operator $\hat p$ acting on the basis
states $|\mu\rangle$ is
\be 
\hat p|\mu\rangle=\frac{\kappa\gamma\hbar|\mu|}{6}|\mu\rangle~,
\ee 
where $\mu$ (a real number) stands for the eigenstates of $\hat p$,
satisfying the orthonormality relation
\be
\langle\mu_1|\mu_2\rangle=\delta_{\mu_1,\mu_2}~.
\ee
The action of the $\widehat{\exp \left[ \frac{i\mu_0}{2} c\right]} $ operator
    acting on basis states $|\mu\rangle$ is
\be
\widehat{\exp \left[ \frac{i\mu_0}{2} c \right]} | \mu \rangle = 
\exp \left[ \mu_0 \frac{{\rm d}}{{\rm d} \mu}
\right] | \mu \rangle = | \mu+\mu_0 \rangle~,
\ee
where $\mu_0$ is any real number.  Thus, in the old quantisation, the operator
$e^{i\mu_0c/2}$ acts as a simple shift operator.

Using the volume operator $\hat V=|\hat p|^{3/2}$, representing the
volume of the elementary cell with eigenvalues
$V_\mu=(\kappa\gamma\hbar|\mu|/6)^{3/2}$, we get\footnote{Being
  concerned with the large scale behaviour of the LQC equations, we
  neglect the sign ambiguity.} ~\cite{aps}
\be
\hat V|\mu\rangle=\left(\frac{\kappa\gamma\hbar|\mu|}{6}\right)^{3/2}
|\mu\rangle~.
\ee
To define the inverse volume operator one has to trace over SU(2)
valued holonomies. Since there is a freedom in choosing the
irreducible representation to perform the trace, an ambiguity $J$
arises\footnote{Note that $J/2$ stands for the spin of the
  representation. Usually one quantises the gravitational part of the
  Hamiltonian constraint using the fundamental $J=1/2$ representation,
  and the ambiguity is only investigated for the matter part.}.  Let
us use the $J=1/2$ irreducible representation of SU(2). The inverse
volume operator is diagonal in the $|\mu\rangle$ basis and is given
by~\cite{abl}
\be
\label{eq:inv_vol}
\widehat{V^{-1}}| \mu\rangle= \left| \frac{6}{\kappa \gamma \hbar \mu_0}
\left(\{V( \mu+\mu_0)\}^{1/3} -\{V( \mu - \mu_0)\}^{1/3} \right) \right|^3
|\mu\rangle~,
\ee
where $\mu_0$ is proportional to the length of the holonomy. Note that
the regulating length $\mu_0$ is the crucial parameter in the quantum
corrections; $\mu_0$ determines the step-size of the resulting
difference equation. In the above equation, Eq.~(\ref{eq:inv_vol}),
the eigenvalues are bounded and approach zero near the classical
singularity; in the classical case the eigenvalues diverge at the
singularity $\mu=0$. The eigenvalues reach their maximum at a
characteristic scale $\mu_0$, at larger scales they approach the
classical values and at smaller scales they are
suppressed~\cite{Vandersloot PhD}.

As in the full LQG theory, there is no operator corresponding to the
connection. Nevertheless, the action of its holonomy is well-defined.
Let us denote by $\hat{h}_i^{(\mu_0)}$ the holonomy along the edge
parallel to the $i^{\rm th}$ basis vector of length $\mu_0 V_0^{1/3}$
with respect to the fiducial metric. Its action on the basis states is
given by~\cite{Ashtekar:2006wn}
\be
\label{eq:hol1}
\hat{h}_i^{(\mu_0)}| \mu \rangle = \left(\widehat{\rm cs} {\bf 1} 
+ 2 \widehat{\rm sn}\tau_{\rm i} \right) | \mu
\rangle~,
\ee
where,
\beq
\label{defin}
 \widehat{\rm cs} |\mu\rangle \equiv \widehat{\cos (\mu_0 c/2)} | \mu
\rangle &=& \left[ \ | \mu+\mu_0\rangle + | \mu -\mu_0\rangle \
\right]/2~, \nonumber \\ \widehat{\rm sn} |\mu\rangle \equiv
\widehat{\sin (\mu_0 c/2)} | \mu \rangle &=& \left[ \ |
\mu+\mu_0\rangle - | \mu -\mu_0\rangle \ \right]/(2i)~.  \eeq
Thus,
\begin{eqnarray}
\label{eq1}
&& \hat{h}_i^{(\mu_0)} \hat{h}_j^{(\mu_0)}\hat{h}^{(\mu_0)-1}_i
\hat{h}^{(\mu_0)-1}_j|\mu\rangle \nonumber\\ &&~~= \left[ \left(
\widehat{\rm cs}^4 - \widehat{\rm sn}^4 \right) {\bf 1} + 2\left(
{\bf 1} - 4 \tau_j \tau_i\right) \widehat{\rm cs}^2\widehat{\rm
sn}^2 +4\left( \tau_i - \tau_i \right) {\bf 1}\widehat{\rm cs}\
\widehat{\rm sn}^3 \right] | \mu \rangle~,
\end{eqnarray}
and
\begin{eqnarray}
\label{eq2}
&&\hat{h}_i^{(\mu_0)} \left[ \hat{h}_i^{(\mu_0)-1}, \hat{V} \right] |
\mu \rangle \nonumber\\ &&~~~~~~= \left( \hat{V}- \widehat{\rm
cs}\hat{V}\widehat{\rm cs} - \widehat{\rm sn}\hat{V}\widehat{\rm sn}
\right) {\bf 1} |\mu \rangle + 2\tau_i \left( \widehat{\rm
cs}\hat{V}\widehat{\rm sn} - \widehat{\rm sn}\hat{V} \widehat{\rm
cs} \right) | \mu \rangle~.
\end{eqnarray}
The gravitational part of the Hamiltonian operator in terms of SU(2)
holonomies and the triad component, in the irreducible $J=1/2$
representation\footnote{With this choice, the Hamiltonian constraint
  is free of the ill-behaving spurious solutions.} ,
reads~\cite{Vandersloot PhD,Ashtekar:2006wn}
\be
\label{eq:ham_g1} 
\hat{\mathcal C}_{\rm grav} = \frac{2i}{\kappa^2 \hbar \gamma^3
  \mu_0^3}{\rm tr} \sum_{ i j k} \epsilon^{ijk} \left(
\hat{h}_i^{(\mu_0)}\hat{h}_j^{(\mu_0)}\hat{h}i^{(\mu_0)-1}
\hat{h}_j^{(\mu_0)-1} \hat{h}_k^{(\mu_0)} \left[
  \hat{h}_k^{(\mu_0)-1},\hat{V} \right]\right)~.
\ee 
As in the full LQG theory, curvature is defined in LQC in terms of
holonomies around closed loops.  This implies that the limit
$\mu_0\rightarrow 0$ of the above operator does not exist, since it
would mean that the area enclosed by loops should be shrunk to
zero. In the underlying quantum geometry, the eigenvalues of the area
operator are discrete, implying that there is a smallest nonzero
eigenvalue, an {\sl area gap} $\Delta$~\cite{al}. This is indeed the
reason for which the WDW differential equation gets replaced by a
difference equation whose step size is controlled by $\Delta$.  Since
$\mu_0$ enters through the holonomies, its value in the fixed lattice
case was fixed by demanding that the eigenvalue of the area operator
be the area gap: $\Delta=2\sqrt{3}\pi\gamma l_{\rm Pl}^2$, implying
$\mu_0=3\sqrt{3}/2$.

The action of the self-adjoint Hamiltonian constraint operator,
$\hat{\mathcal H}_{\rm grav}=(\hat{\mathcal C}_{\rm grav} + \hat{\mathcal
C}_{\rm grav}^{\dagger} )/2$, on the basis states, $|\mu\rangle$, is
\be
\hat{\mathcal H}_{\rm grav} | \mu \rangle 
=
\frac{3}{4\kappa^2 \gamma^3 \hbar \mu_0^3}
\Bigl\{ \bigl[ R(\mu)+R(\mu+4\mu_0)\bigr]| \mu+4\mu_0 \rangle
-4R(\mu ) | \mu\rangle +\bigl[ R(\mu)+R(\mu-4\mu_0)\bigr]| 
\mu-4\mu_0 \rangle \Bigr\}~,
\ee
where 
\be
R(\mu )= \left( \kappa \gamma \hbar/6\right)^{3/2} 
\Big| | \mu+\mu_0 |^{3/2} -|\mu-\mu_0|^{3/2} \Big|~.
\ee

Having the Hamiltonian operator, dynamics are determined by the
Hamiltonian constraint\footnote{Note that the Gauss and the
  diffeomorphism constraints are automatically satisfied by
  appropriate gauge fixing.}
\be
\left(\hat{\mathcal H}_{\rm grav} +\hat{\mathcal H}_\phi\right) |\Psi\rangle=0~.
\ee
Note that in the full LQG theory, there is an infinite number of
constraints, whereas in the reduced homogeneous and isotropic case
there is only one integrated Hamiltonian constraint.  Matter is then
introduced by just adding the actions of matter components to the
gravitational action. One finally obtains difference equations
analogous to the differential WDW equations.\footnote{The reader
  should note that in LQC the physical fundamental object is the
  discrete difference equation, while the differential equation is
  just the approximation in the continuum limit.}

Let us impose the constraint equation on the physical wave-functions
$|\Psi\rangle$, which can be expanded using
the basis states as $|\Psi\rangle = \sum_\mu
\Psi_\mu(\phi)|\mu\rangle$, with summation over values of $\mu$ and
where the dependence of the coefficients on $\phi$ represents the
matter degrees of freedom. Since the states $|\mu\rangle$ are
eigenstates of the triad operator, the coefficients $\Psi_\mu(\phi)$
represent the state in the triad representation.  Thus, quantising the
Friedmann equation along the lines of the constraint in the full LQG
theory, one gets the following difference
equation~\cite{Vandersloot:2005kh}
\beq\label{qee}
&&~~\Biggl[\Big| V_{\mu+5\mu_0}-V_{\mu+3\mu_0}\Big|+\Big|V_{\mu+\mu_0}
 - V_{\mu-\mu_0}\Big|\Biggr] \Psi_{\mu+4\mu_0}(\phi) -
 4\Big|V_{\mu+\mu_0}V_{\mu-\mu_0}\Big| \Psi_\mu(\phi) \nonumber \\
 &&+\Biggl[\Big|V_{\mu-3\mu_0}-
 V_{\mu-5\mu_0}\Big|+\Big|V_{\mu+\mu_0}- V_{\mu-\mu_0}\Big|\Biggr]
 \Psi_{\mu-4\mu_0}(\phi) =- \frac{4\kappa^2 \gamma^3 \hbar \mu_0^3}{3}
 {\mathcal H}_{\rm \phi}(\mu)\Psi_\mu(\phi)~,
\eeq
where the matter Hamiltonian $\hat{\mathcal H}_{\rm \phi}$ is assumed
to act diagonally on the basis states with eigenvalue ${\mathcal
H}_{\rm \phi}(\mu)$. Equation (\ref{qee}) is indeed the quantum
evolution (in internal time $\mu$) equation. There is no continuous
variable (the scale factor in classical cosmology), but a label $\mu$
with discrete steps.  The wave-function $\Psi_\mu(\phi)$, depending on
internal time $\mu$ and matter fields $\phi$, determines the dependence
of matter fields on the evolution of the universe.
A massless scalar field plays the r\^ole of the {\sl emergent time}.
Thus, in LQC the quantum evolution is governed by a second order
difference equation, rather than the second order differential equation
of the WDW quantum cosmology. As the universe becomes large and enters
the semi-classical regime, the WDW differential equation becomes a
very good approximation to the difference equation of LQC.

\section{Lattice refinement}

Consider the continuum limit (namely that $\mu\gg\mu_0$) of the
Hamiltonian constraint operator acting on the physical states. In the
small regulating length $\mu_0$ limit, one obtains~\cite{ns2} a
second order difference equation which distinguishes the components of
the wave-functions in different lattices of spacing $4\mu_0$.
Assuming that $\Psi$ does not vary much on scales of the order of
$4\mu_0$, known as {\sl pre-classicality}~\cite{pre-cl}, one can
smoothly interpolate between the points on the discrete function
$\Psi_\mu(\phi)$ and approximate them by the continuous function
$\Psi(\mu,\phi)$. In this way, one approximates the difference
equation by a differential equation for a continuous wave-function.

The form of the wave-functions indicates that the period of
oscillations can decrease as the scale increases, which implies that
at sufficiently large scales the assumption of {\sl pre-classicality}
breaks down\footnote{The constant lattice, in the {\sl old
    quantisation} approach, does not take into account the expansion
  of the fiducial cell in a FLRW
  background.}. This would then lead to quantum gravity corrections
at large scales (i.e., classical) physics. To avoid this undesired
event, was one of the motivations behind lattice
refinement~\cite{Vandersloot:2005kh,lr1}.  Allowing the length scale
of the holonomies to vary, the form of the difference equation
changes. Assuming that the lattice size is growing, the step-size of
the difference equation is not constant in the original triad
variables. The exact form of the difference equation depends on the
lattice refinement used.

Consider the particular model
\be
\mu_0\rightarrow\tilde\mu(\mu)=\mu_0\mu^{-1/2}~;
\ee
we will come back to the issue of the lattice refinement choice in a
subsequent section.

The basic operators are given by replacing $\mu_0$ with $\tilde{\mu}$.
Upon quantisation~\cite{Ashtekar:2006wn}
\be
\widehat{ e^{i\tilde{\mu}c/2}}
|\mu \rangle=e^{-i\tilde{\mu}\frac{\rm d}{{\rm d}\mu} }|\mu\rangle~,
\ee
which is no longer a simple shift operator since $\tilde{\mu}$ is a 
function of $\mu$. Changing the basis to 
\be\label{eq:basis}
 \nu = \mu_0 \int \frac{{\rm d}\mu}{\tilde{\mu}}=\frac{2}{3} \mu^{3/2}~,
\ee
one gets
\be
 e^{-i\tilde{\mu}\frac{\rm d}{{\rm d} \mu}}|\nu\rangle 
= e^{-i\mu_0 \frac{\rm d}{
{\rm d}\nu}}|\nu\rangle = |\nu+\mu_0\rangle~.
\ee
The volume operator acts on these basis states as
\be
 \hat{V}|\nu\rangle 
= \frac{3\nu}{2}\left(\frac{\kappa\gamma\hbar}{6}\right)^{3/2}
|\nu\rangle~,
\ee
and the self-adjoint Hamiltonian constraint operator acts
as~\cite{Vandersloot PhD}
\begin{eqnarray}
 &&{\hskip-.6truecm}\hat{\mathcal H}_{\rm g} |\nu\rangle =
  \frac{9|\nu|}{16\mu_0^3} \left(
  \frac{\hbar}{6\kappa\gamma^3}\right)^{1/2}\nonumber\\ 
&&{\hskip-.6truecm}\times
  \Biggl[ \frac{1}{2}\Bigl\{ U\left( \nu \right) +U\left( \nu+4\mu_0
    \right) \Bigr\} |\nu+ 4\mu_0\rangle -2U\left( \nu \right) |\nu
    \rangle + \frac{1}{2} \Bigl\{ U\left( \nu \right) + U\left(
    \nu-4\mu_0\right) \Bigr\} |\nu - 4 \mu_0 \rangle \Biggr],
\end{eqnarray}
where 
\be
U\left(\nu\right) = |\nu + \mu_0 | - | \nu - \mu_0|~.
\ee
Expanding $|\Psi\rangle=\sum_\nu \Psi_\nu(\phi) |\nu\rangle$ the
Hamiltonian constraint reads~\cite{ns2}
\beq\label{eq:const_refine}
&& \ \  \frac{1}{2} \Big|\nu + 4\mu_0 \Big|
\Bigl[ U\left(\nu + 4\mu_0\right) +U\left( \nu\right) \Bigr]  
\Psi_{\nu+4\mu_0}\left(
\phi \right) 
+ 2|\nu| U\left(\nu\right) \Psi_\nu \left(\nu\right) \nonumber\\
&& +\frac{1}{2} \Big|\nu-4\mu_0\Big| \Bigl[ U\left(\nu-4\mu_0\right) +
 U\left( \nu\right)
\Bigr] \Psi_{\nu-4\mu_0}\left( \phi \right) \nonumber \\
&& = 
- \frac{16\mu_0^3}{9}\left(\frac{6\kappa\gamma^3}{\hbar}\right)^{1/2}
 {\mathcal H}_{\phi}\left(\nu\right)
 \Psi_{\nu} \left( \phi\right)~;
\eeq
$ {\mathcal H}_{\phi}$ stands for the matter part of the
Hamiltonian, which for a massive scalar field is given by
\be
 {\mathcal H}_{\phi}=\kappa\left[\frac{P_\phi^2}{2a^3} + a^3V(\phi)\right]~,
\ee
with $P_\phi$ the momentum and $V(\phi)$ the potential of the scalar
field $\phi$. Quantising the Hamiltonian constraint we obtain, in terms
of $p=\kappa\gamma\hbar\mu/6$,
\be
\sqrt{p}\frac{\partial^2 \Psi(p,\phi)}{\partial p^2}
+\frac{\partial^2}{\partial p^2} \left(\sqrt{p}\Psi(p,\phi)\right) - 3
p^{-3/2} \frac{\partial^2 \Psi(p,\phi)}{\partial \phi^2}
+\frac{6}{\kappa \hbar^2}p^{3/2} V(\phi) \Psi(p,\phi) +{\mathcal
O}\left(\mu_0\right)+\cdots = 0~, 
\ee
which is a particular factor ordering of the WDW equation for a
massive scalar field.

\subsection{Implications for inflation}

We will show that lattice refinement is essential in order to achieve
a successful inflationary era. Let us consider a fixed and a
dynamically varying lattice and solve in the continuum limit the
second order difference equation which governs the quantum
evolution. By doing so we constrain the potential of a scalar field
(the inflaton) so that the continuum approximation is valid. A second
constraint is imposed on the inflationary potential so that there is
consistency with measurements of the cosmic microwave background
temperature anisotropies on large angular scales. Combining the two
constraints in a particular inflationary model, for either a fixed
lattice or a given lattice refinement model, we deduce the conditions
for natural and successful inflation within LQC in each of the two
cases.

More precisely, let us separate the wave-function $\Psi(p,\phi)$ into
$\Psi(p,\phi)= \Upsilon(p)\Phi(\phi)$ and approximate the dynamics of
the inflaton field, $\phi$, by setting $V(\phi) =V_\phi
p^{\delta-3/2}$, where $V_\phi$ is a constant and $\delta=3/2$ in the
case of slow-roll, to get~\cite{ns2}
\be\label{eq-p} 
p^{-1/2}\frac{{\rm d}}{{\rm d} p} \left[ p^{-1/2} \frac{{\rm d}}{{\rm
      d} p } \left( p^{3/2} \Upsilon\left(p\right) \right) \right] +
\beta V_\phi p^\delta\Upsilon(p) =0~,
\ee
with solutions~\cite{ns2}
\be
 \Upsilon\left( p\right) \approx  p^{-(9+2\delta)/8} \sqrt{\frac{
2\delta+3}{2\sqrt{\beta V_\phi}\pi} }
\Biggl[ C_1 \cos \left( x-\frac{3\pi}{2(2\delta+3)}
-\frac{\pi}{4} \right)
+C_2 \sin \left( x -\frac{3\pi}{2(2\delta+3)}-\frac{\pi}{4} \right)
\Biggr]~,
\ee
where 
\be 
x=4\sqrt{\beta V_\phi}(2\delta+3)^{-1} p^{(2\delta+3)/4}~,
\ee 
and $\beta=96/(\kappa\hbar^2)$.  

Without lattice refinement, the discrete nature of the underlying
lattice would eventually be unable to support the oscillations and the
assumption of pre-classicality will break down, implying that the
discrete nature of space-time becomes significant on very large
scales. For the end of inflation to be describable using classical GR,
it must end before a scale, at which the assumption of
pre-classicality breaks down and the semi-classical description is no
longer valid, is reached. We will quantify this constraint.

The separation between two successive zeros of $ \Upsilon\left(
p\right) $ is
\be
 \Delta p =\frac{\pi}{\sqrt{\beta V_\phi}} p^{(1-2\delta)/4}~.  
\ee
For the continuum limit to be valid, the wave-function must vary
slowly on scales of the order of $\mu_c = 4\tilde{\mu}$.
Thus, we impose the constraint~\cite{ns2}
\be
\Delta p > 4\mu_0\left(\frac{\kappa\gamma\hbar}{6}\right)^{3/2}p^{-1/2}~,
\ee
which implies the following constraint on $V_\phi$~\cite{ns2}
\be
 V_\phi < \frac{27\pi^2}{192\mu_0^2\gamma^3 \kappa^2\hbar} 
p^{(3-2\delta)/2}~.
\ee
For slow-roll inflation, $V(\phi)$ must be approximately constant,
implying $\delta\approx 3/2$.  For $\mu_0=3\sqrt{3}/2$ and
$\gamma\approx 0.24$, the constraint on the inflationary
potential in units of $\hbar=1$ reads 
\be 
V(\phi) \lsim 2.35\times10^{-2} l_{\rm Pl}^{-4}~.
\ee 
This is a softer constraint than the one imposed for fixed lattices,
namely~\cite{ns2}
\be
V_\phi \ll 10^{-28}l_{\rm pl}^{-4}~,
\ee
assuming that half of inflation takes place during the classical era.

Selecting a successful and simple inflationary model, for example
$V(\phi)=m^2\phi^2/2$, an additional constraint can be imposed on the
potential so that the fractional over-density in Fourier space at
horizon crossing is consistent with the COBE-DMR
measurements. Combining the two constraints we obtain~\cite{ns2}
for the fixed and varying lattices
\begin{eqnarray}
\label{constraint_on_m}
 m &\lsim&  70 (e^{-2N_{\rm cl}}) ~M_{\rm Pl} 
\\ 
\mbox {and}~~ m &\lsim& 10 ~M_{\rm Pl}~,
\end{eqnarray}
respectively.

Thus, for any significant proportion of inflation to take place during
the classical era, the constraint imposed on the inflaton mass is very
strong, while it becomes natural once lattice refinement is taken into
account.

\subsection{Relation between lattice refinement model and matter Hamiltonian}

A particular lattice refinement model can support only certain types
of matter~\cite{Nelson:2007um}.  To prove it, let us parametrise the
lattice refinement by a parameter $A$ and the matter Hamiltonian by a
parameter $\delta$ and solve the Hamiltonian constraint. The
restrictions on the two-dimensional parameter space will become
apparent once we impose some physical restrictions to the solutions of
the wave-functions~\cite{Nelson:2007um}.

More precisely, assuming
\be\tilde{\mu}=\mu_0 \mu^A~,
\ee
we obtain
\be
\label{def:nu}
\nu=\frac{\tilde\mu_0\mu^{1-A}}{\mu_0(1-A)}~.
\ee 
The WDW constraint equation reads
\be 
\left(\hat{\cal H}_{\rm grav} +\hat{\cal H}_\phi\right)\Psi = 0~.  
\ee
Being interested in the large scale limit, we can
approximate~\cite{Nelson:2007um} the matter Hamiltonian with
$\hat{\cal H}_\phi= \hat{\nu}^\delta\hat{\epsilon}\left(\phi\right)$,
leading to 
\be 
\hat{\epsilon}\left(\phi\right)\Psi \equiv
\epsilon\left(\phi\right)\Psi = -\nu^{-\delta}\hat{\cal H}_{\rm grav}
\Psi~.  
\ee
A necessary condition for the wave-functions to be physical
is that the finite norm of the physical wave-functions, which is defined by
$\int_{\phi=\phi_0}d\nu
|\nu |^\delta \overline{\Psi}_1\Psi_2$, must 
be independent of the choice of
$\phi=\phi_0$. The solutions of the constraint are
renormalisable provided they decay, on large scales, faster than
$\nu^{-1/(2\delta)}$.

To solve the constraint equation, we need to specify the from of
${\cal H}_\phi$, which has in general two terms with different scale
dependence. Being interested in the large scale limit, one will be the
dominant term, allowing to write~\cite{Nelson:2007um}
\be
\beta{\mathcal H}_{\phi} =\epsilon_\nu(\phi) \nu^{\delta_\nu}~,
\ee
where the function $\epsilon_\nu$ is constant with respect to $\nu$.

We then solve~\cite{Nelson:2007um} the constraint equation and
consider only those solutions which are the physical ones.  The
large scale behaviour of the wave-functions must be normalisable --- a
necessary condition for having physical wave-functions --- and the
wave-functions should preserve pre-classicality at large scales --- a
necessary condition for the validity of the continuum
limit~\cite{Nelson:2007um}. This procedure leads to constraints to the
two-dimensional parameter space $(A,\delta)$, which we illustrate
below~\cite{Nelson:2007um} in Fig.~1 for $A$ in the range $0<A<-1/2$,
imposed from full LQG theory considerations~\cite{Bojowald:2007ra}.

\begin{figure}
 \begin{center}
  \input{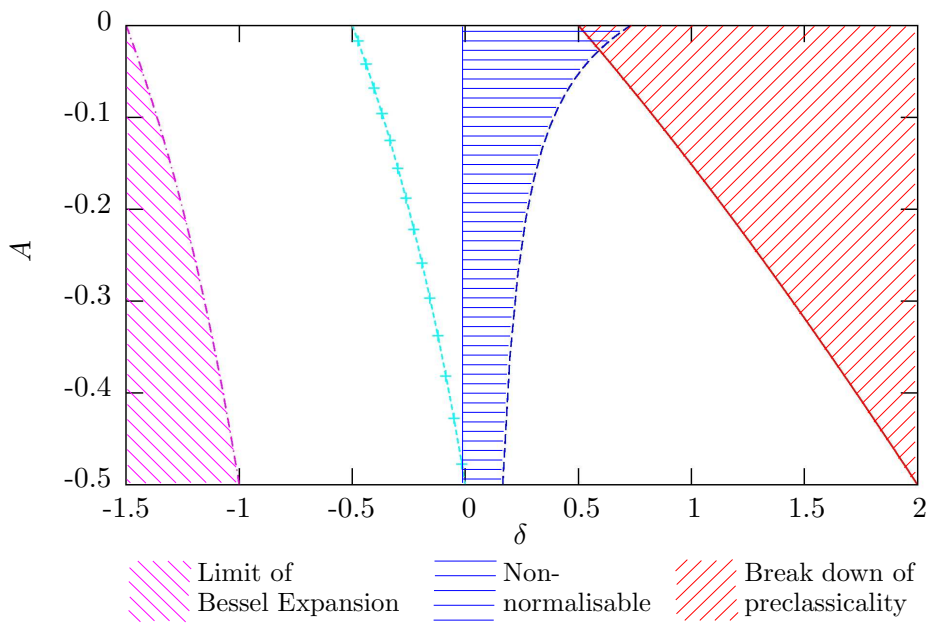}
  \caption{\label{fig6} The allowed types of matter content are
    significantly restricted. Note that in the case of a varying
    lattice ($A \neq 0$) it is not always possible to treat the large
    scale behaviour of the wave-functions perturbatively (dashed line
    with crosses)~\cite{Nelson:2007um}.}
 \end{center}
\end{figure}

We thus conclude that the continuum limit of the Hamiltonian
constraint equation is sensitive to the choice of model and only a
limited range of matter components can be supported within a
particular choice~\cite{Nelson:2007um}.

\subsection{Numerical methods for solving the constraint equation for any 
lattice refinement model}

The lattice refinement implies new dynamical difference equations,
which are not expected to have a uniform step-size, leading to
technical complications. This becomes apparent in the case of
two-dimensional wave-functions, such as those necessary to study
Bianchi models of black hole interiors. More precisely, the
information needed to calculate the wave-function at a given lattice
point is not provided by previous iterations. We prescribe below a
method~\cite{ns3} based on Taylor expansion that can be used to
perform this desired interpolations with a well-defined and
predictable accuracy\footnote{A simple local interpolation scheme to
  approximate the necessary data points, allowing direct numerical
  evolution of two-dimensional systems has been also proposed in
  Ref.~\cite{nt1}.}.

Let us first note that the Hamiltonian constraint for a
one-dimensional difference equation defined on a varying lattice, can
be mapped onto a fixed lattice simply by a change of
basis~\cite{ns3}\footnote{We have checked~\cite{ns3} the validity of
  our Taylor expansion numerical method in calculating the
  wave-function by comparing our results with those obtained by
  mapping the one-dimensional difference equation defined on a varying
  lattice, onto a fixed lattice by performing a change of basis.}.
However this method is of no help for the two-dimensional case, where
the Hamiltonian constraint is a difference equation on a varying
lattice~\cite{Bojowald:2007ra},

\beq
\label{eq:vary2D}
&&C_{+}\left(\mu,\tau\right) \left[
  \Psi_{\mu+2\delta_\mu,\tau+2\delta_\tau} - \Psi_{\mu-2\delta_\mu,
    \tau+2\delta_\tau} \right] \nonumber \\ 
&&+ C_0\left(\mu,\tau\right) \left[
  \left(\mu+2\delta_\mu\right)\Psi_{\mu+4\delta_\mu, \tau} - 2\left( 1
  + 2 \gamma^2 \delta_\mu^2\right) \mu\Psi_{\mu,\tau}
  +\left(\mu-2\delta_\mu\right)\Psi_{\mu-4\delta_\mu,\tau} \right]
\nonumber \\ 
&&+C_{-} \left( \mu,\tau\right) \left[
  \Psi_{\mu-2\delta_\mu,\tau-2\delta_\tau} -
  \Psi_{\mu+2\delta_\mu,\tau-2\delta_\tau}\right] = \frac{\delta_\tau
  \delta_\mu^2}{\delta^3} {\cal H}_\phi\Psi_{\mu,\tau}~, 
\eeq 
with
\beq 
C_{\pm} &\equiv& 2\delta_\mu \left( \sqrt{\left| \tau \pm 2
  \delta_\tau \right| } + \sqrt{\left| \tau\right| } \right)~, \\ 
C_0 &\equiv& \sqrt{\left| \tau + \delta_\tau\right|} - \sqrt{ \left| \tau -
  \delta_\tau\right| }~, 
\eeq 
where we have defined $\delta_\mu$ and $\delta_\tau$ as the step-sizes
along the $\mu$ and $\tau$ directions, respectively.  The parameter
$\delta$, with $0<\delta<1$, gives the fraction of a lattice edge that
the underlying graph changing Hamiltonian
uses~\cite{Bojowald:2007ra}.

In the case of lattice refining, $\delta_\mu$ and $\delta_\tau$
are decreasing functions of $\mu$ and $\tau$, respectively, and the data
needed to calculate the value of the wave-function at a particular
lattice site are not given by previous iterations. One can use Taylor
expansions to calculate the necessary data points~\cite{ns3}.  More
precisely, let us assume that the matter Hamiltonian acts diagonally on the
basis states of the wave-function, namely
\be
\hat{\cal H}_\phi |\Psi \rangle \equiv \hat{\cal H}_\phi
\sum_{\mu,\tau} \Psi_{\mu,\tau} |\mu,\tau\rangle = \sum_{\mu,\tau}
    {\cal H}_\phi \Psi_{\mu,\tau} |\mu,\tau \rangle~.  
\ee
Given a function evaluated at three (noncolinear) coordinates, the
Taylor approximation to the value at a fourth position is 
\beq
 f\left( x_4,y_4\right) &=& f\left(x_2, y_2\right) + \delta^x_{42}
 \frac{\partial f}{\partial x} \Big|_{x_2, y_2} + \delta^y_{42}
 \frac{\partial f}{\partial y} \Big|_{x_2, y_2} \nonumber\\ 
&& + {\cal O}\left( \left(\delta^x_{42}\right)^2 \frac{\partial^2
   f}{\partial x^2}\Big|_{x_2,y_2}\right)+{\cal O}\left(
 \left(\delta^y_{42}\right)^2 \frac{\partial^2 f}{\partial
   y^2}\Big|_{x_2,y_2}\right) ~,\label{eq:taylor} 
\eeq 
where the Taylor expansion is taken about the position
$\left(x_2,y_2\right)$, we have defined $\delta_{ij}^x \equiv x_i-x_j$ and
$\delta_{ij}^y \equiv y_i-y_j$, and the differentials can be approximated using
\beq
f\left(x_1,y_1\right) &=& f\left(x_2,y_2\right) + \delta^x_{12} \frac{
  \partial f}{\partial x}\Big|_{ x_2,y_2} + \delta^y_{12}
\frac{\partial f}{\partial y}\Big|_{x_2,y_2} + \cdots~,\\
f\left(x_3,y_3\right) &=& f\left(x_2,y_2\right) + \delta^x_{32} \frac{
  \partial f}{\partial x}\Big|_{ x_2,y_2} + \delta^y_{32}
\frac{\partial f}{\partial y}\Big|_{x_2,y_2} + \cdots~,
\eeq
where the dots indicate higher order terms.

For slowly varying wave-functions, it has been shown~\cite{ns3} that
linear approximation is very accurate and higher order terms in Taylor
expansion can only improve the accuracy by $10^{-2}\%$.  This method
can be applied in any lattice refinement model, while its accuracy can
be estimated. Even though we have illustrated it in the case of black
hole interiors, this method can be applied to anisotropic Bianchi
models and in general to systems with anisotropic symmetries. 

By using this Taylor expansions method, we were able to
confirm~\cite{ns3} numerically the stability criterion of the
Schwarzchild interior, which was earlier found~\cite{Bojowald:2007ra}
using a von Neumann analysis, and investigate~\cite{ns3} how lattice
refinement can change the stability properties of the system. Finally,
the underlying discreteness of space-time leads to a twist~\cite{ns3}
in the wave-functions, for both a constant lattice, as well as lattice
refinement models.

\subsection{Uniqueness of WDW factor ordering and the lattice refinement 
choice} 

The {\sl correct} lattice refinement model should in principle be
given by the full LQG theory. In this sense, one should consider the
full Hamiltonian constraint and find the way that its action balances
the creation of new vertices while the volume increases.  Instead of
doing so, as we have already discussed earlier, phenomenological
arguments~\cite{Nelson:2007um, ns2} have been used to constrain the
choice of the lattice refinement model by the form of the matter
Hamiltonian. Later on, it has been argued~\cite{gs08} that only the
lattice refinement model $\tilde\mu=\mu_0\mu^{-1/2}$~\cite{ns4}, can be
achieved by physical considerations of large scale physics and
consistency of the quantisation structure. We will show below that
this choice is also the only one which makes the factor order
ambiguities of LQC to disappear in the continuum limit~\cite{ns4}.

Indeed, there are many ways of writing the gravitational part of the
Hamiltonian constraint in terms of the triad and the holonomies of the
connection, our quantisable variables.
Writing~\cite{Ashtekar:2006wn} for example,
\be\label{eq:ham} 
\hat{\mathcal C}_{\rm grav} = \frac{2i \ }{\kappa^2 \hbar \gamma^3
  k^3} {\rm tr} \sum_{\rm ijk} \epsilon^{\rm ijk} \left( \hat{h}_{\rm
  i} \hat{h}_{\rm j} \hat{h}_{\rm i}^{-1} \hat{h}_{\rm
  j}^{-1}\hat{h}_{\rm k} \left[ \hat{h}_{\rm k}^{-1},\hat{V} \right]
\right)~,
\ee
one immediately realises that there are many possible choices of
factor ordering that could have been made at this point, since
classically the actions of the holonomies commute.  However, each of
these factor ordering choices leads to a different factor ordering of
the WDW equation in the continuum limit.

The action of the factor ordering chosen in Eq.~(\ref{eq:ham}) above
gives~\cite{ns4}
\be\label{eq:factor1}
\epsilon_{\rm ijk} {\rm tr}  \left( \hat{h}_{\rm i} \hat{h}_{\rm j}
\hat{h}_{\rm i}^{-1} \hat{h}_{\rm j}^{-1}\hat{h}_{\rm k} \left[ 
\hat{h}_{\rm k}^{-1},\hat{V} \right] \right)  =  
-24\widehat{\rm sn}^2\widehat{\rm cs}^2 \left( \widehat{\rm cs} \hat{V}
\widehat{\rm sn} - \widehat{\rm sn} \hat{V} \widehat{\rm cs}\right)~,
\ee
while other choices have different action, as it has been explicitly
found in Ref.~\cite{ns4}.

Defining $\hat{V}|\nu\rangle = V_{\nu}|\nu\rangle$, the action of the
above factor ordering on a general state in the Hilbert space given by
$|\Psi \rangle = \sum_\nu \psi_\nu|\nu\rangle$ reads~\cite{ns4}
\beq\label{eq:ham1.2}
\epsilon_{\rm ijk} {\rm tr}  \left( \hat{h}_{\rm i} \hat{h}_{\rm j}
\hat{h}_{\rm i}^{-1} \hat{h}_{\rm j}^{-1}\hat{h}_{\rm k} \left[ 
\hat{h}_{\rm k}^{-1},\hat{V} \right] \right)|\Psi\rangle & = & 
\frac{-3i}{4} \sum_\nu \Biggl[ \Bigl( V_{\nu-3k} - V_{\nu-5k} \Bigr) 
\psi_{\nu-4k}
-2\Bigl(V_{\nu+k}- V_{\nu-k}\Bigr)\psi_\nu \Biggr. \nonumber \\
&&+\Biggl. \Bigl( V_{\nu+5k} - V_{\nu+3k} \Bigr) \psi_{\nu+4k} \Biggr] |
\nu\rangle~.
\eeq
Similarly, one can find~\cite{ns4} the action of any other factor
ordering choice.

Then one can take the continuum limit of these expressions by
expanding $\psi_\nu \approx \psi\left(\nu\right)$ as a Taylor
expansion in small $k/\nu$. For the factor ordering
chosen for this illustration here, the large scale continuum limit of
the Hamiltonian constraint reads~\cite{ns4}:
\beq\label{eq:final}
\lim_{k/\nu \rightarrow 0} 
\epsilon_{\rm ijk} {\rm tr}  \left( \hat{h}_{\rm i} \hat{h}_{\rm j}
\hat{h}_{\rm i}^{-1} \hat{h}_{\rm j}^{-1}\hat{h}_{\rm k} \left[ 
\hat{h}_{\rm k}^{-1},\hat{V} \right] \right)|\Psi\rangle \sim  
\nonumber \\
\frac{-36i}{1-A} \alpha^{3/\left[2\left(1-A\right)\right]} 
k^3 \sum_\nu \nu^{\left(1+2A\right)/\left[2\left(
1-A\right)\right]}
 \Biggl[ \frac{{\rm d}^2 \psi}{{\rm d} \nu^2 } + 
\frac{1+2A}{1-A} \frac{1}{\nu} \frac{{\rm d}
\psi}{{\rm d}\nu} + \frac{\left(1+2A\right)
\left(4A-1\right)}{\left(1-A\right)^2} \frac{1}{4\nu^2}
\psi\left(\nu\right)
\Biggr] |\nu\rangle~. \nonumber \\ 
\eeq
Setting $\alpha=3\mu_0/(2k)$ and $\mu_0=k$, all lattice refinement
models will lead to the same continuum limit for the WDW equation,
only for $A=-1/2$~\cite{ns4}, in which case the WDW equation
reads~\cite{ns4} \be\label{eq:final_con}
\lim_{k/\nu \rightarrow 0} {\cal C}_{\rm grav} |\Psi\rangle = 
\frac{72}{\kappa^2 \hbar \gamma^3}
\left(\frac{\kappa\gamma\hbar}{6}\right)^{3/2} \sum_\nu \frac{{\rm d}^2
\psi}{{\rm d} \nu^2} |\nu\rangle~.
\ee
Thus, there is only one lattice refinement model, namely
$\tilde\mu=\mu_0\mu^{-1/2}$, with a non ambiguous continuum
limit\footnote{ The $A=-1/2$ choice can be easily understood with the
  following simple argument~\cite{ashtekar-priv}. In LQC, the basis
  sates are $|\mu\rangle$ and the physical area of the fiducial cell
  is $\mu V_0^{2/3}$.  Consider, in the full LQG theory, $N$ fluxes
  passing through a side of the fiducial cell and divide its surface
  in $\Delta$ elementary surfaces. Then $N\Delta=\mu V_0^{2/3}$,
  implying $N=\mu V_0^{2/3}/\Delta$. The holonomies of the connection
  are $e^{i\lambda c/2}=e^{i\lambda\tilde cV_0^{1/3}/2}$ and the
  fiducial area is $\lambda^2=V_0^{2/3}/N$. Thus, $\lambda\propto
  \mu^{-1/2}$.}.

In conclusion, phenomenological and consistency requirements lead to a
particular lattice refinement model, implying that LQC predicts a
unique factor ordering of the WDW equation in its continuum limit.
Alternatively, demanding that factor ordering ambiguities disappear
in LQC at the level of WDW equation leads to a unique choice for the
lattice refinement model.

\section{Conclusions}

LQG canonically quantises space-time via triad and holonomies of the
connection. Full understanding of the theory has not yet been reached,
nevertheless symmetry reduction versions akin to WDW mini-superspace
model have been successfully developed.

As a first approximation the quantised holonomies were taken to be
shift operators with a fixed magnitude. This results in the quantised
Hamiltonian constraint being a difference equation with a constant
interval between points on the lattice. These models lead to serious
instabilities in the continuum semi-classical limit.

In the underlying LQG theory, the contributions to the discrete
Hamiltonian operator depend on the state which describes the
universe. As the universe expands, the number of contributions
increases, so that the Hamiltonian constraint operator is creating new
vertices of a lattice state, leading to a refinement of the discrete
lattice in LQC.

The lattice refinement can be modelled and the instabilities in the
continuum era get eliminated.  We have discussed here why lattice
refinement seems to be necessary to achieve a natural inflationary
era, and we have illustrated that only a limited range of matter
components can be supported within a particular lattice refinement
choice. We have then shown that factor ordering ambiguities in the
continuum limit of the gravitational part of the Hamiltonian
constraint disappear only for a particular choice of lattice
refinement.

Whilst the continuum limit of the lattice refinement models can be
taken, there is a complication in directly evolving two-dimensional
wave-functions, such as those necessary to study Bianchi models or
black hole interiors. The information needed to calculate the
wave-function at a given lattice point is not provided by previous
iterations.  We have shown that Taylor expansions can be used to
perform this interpolation with a well-defined and predictable
accuracy. We have then discussed how lattice refinement can alter
stability conditions of the system.

We have only focused here in a few aspects of LQC concerning lattice
refinement. There is certainly a much vaster arena of themes within
LQC, an area which is gaining a lot of interest from the scientific
community. Its basic advantage is that it allows us to successfully
address some physical features of our universe, while it gives us some
valuable insight for the full LQG theory.

\ack It is a pleasure to thank the organisers of the NEBXIII
meeting\footnote{http://www.astro.auth.gr/~neb-13/} {\sl Recent
  Developments in Gravity} ---
$N\epsilon\omega\tau\epsilon\rho\epsilon$s
$E\xi\epsilon\lambda\iota\xi\epsilon\iota$s $\sigma\tau\eta$
$B\alpha\rho\upsilon\tau\eta\tau\alpha$ --- for inviting me to give
this talk, in the beautiful city of Thessaloniki in Greece.  This work
is partially supported by the European Union through the Marie Curie
Research and Training Network {\sl UniverseNet} (MRTN-CT-2006-035863).

\end{document}